\documentclass[%
reprint,
superscriptaddress,
 amsmath,amssymb,
 aps,
]{revtex4-2}

\usepackage{blindtext}
\usepackage{graphicx}
\usepackage{dcolumn}
\usepackage{bm}
\usepackage[colorlinks=true,allcolors=blue]{hyperref}%

\usepackage{siunitx}
\usepackage{physics}

\makeatletter
\def\maketitle{
\@author@finish
\title@column\titleblock@produce
\suppressfloats[t]}
\makeatother

\newcommand{\beginsupplement}{%
        \setcounter{table}{0}
        \renewcommand{\thetable}{S\arabic{table}}%
        \setcounter{figure}{0}
        \renewcommand{\thefigure}{S\arabic{figure}}%
}

\usepackage{bibunits}
\defaultbibliographystyle{apsrev4-2}
\defaultbibliography{refs}

\begin{document}
\renewcommand{\appendixname}{}
\newcommand{\hphys}{Department of Physics, Harvard University, Cambridge, Massachusetts, USA}
\newcommand{\heng}{School of Engineering \& Applied Sciences, Harvard University, Cambridge, Massachusetts USA}
\newcommand{\stan}{Department of Physics, Stanford University, Stanford, California, USA}
\def\SIMES{Stanford Institute for Materials and Energy Sciences, SLAC National Accelerator Laboratory, Menlo Park, CA 94025}

\title{Microwave spin resonance in epitaxial thin films of spin liquid candidate $\text{TbInO}_3$}

\author{Sandesh S. Kalantre}
\affiliation{\stan}
\affiliation{\SIMES}

\author{Johanna Nordlander}
\affiliation{\hphys}
\affiliation{Department of Physics, University of Zurich, Zurich, Switzerland}
\author{Margaret A. Anderson}
\affiliation{\hphys}

\author{Julia A. Mundy}
\affiliation{\hphys}
\affiliation{School of Engineering and Applied Sciences, Harvard University, Cambridge, MA, USA}

\author{David Goldhaber-Gordon}
\affiliation{\stan}
\affiliation{\SIMES}

\date{\today}

\begin{abstract}
Minimizing the energy of a many body system tends to favor order, but classical frustration and quantum fluctuations destabilize that order. The tension between these effects can produce exotic quantum states of matter. 
Quantum spin liquid (QSL) states emerge in models of localized magnetic moments where the crystal lattice connectivity frustrates ordering, and the exchange interaction of neighboring spins strengthens quantum fluctuations. Experimentally identifying a QSL in a real material is challenging from the lack of an order parameter. Piecing together evidence from varied techniques is necessary for diagnosing the nature of the ground state -- QSL or otherwise -- of a frustrated spin system.  In this work, we use coplanar superconducting resonators to probe magnetic excitations in epitaxially grown thin films of a spin liquid candidate $\text{TbInO}_3$. 
Adapting microwave techniques from the field of circuit quantum electrodynamics, we measure responses of these thin films whose volume is too low for applying conventional bulk techniques. 
In-plane susceptibility extracted from the spin resonance signal indicates extreme frustration of magnetic order down to $20\,\si{\milli\kelvin}$, over two orders of magnitude lower than the Curie-Weiss energy scale. 
Through a crystal field analysis, we identify the doublet eigenstates comprising the ground state. As a consequence of improper ferroelectricity, Tb moments split into two flavors with distinct g-factors reflecting the local crystal field environment of each site.  
Spin-orbit coupling, crystal fields, magnetic frustration and improper ferroelectricity distinctively combine to shape the magnetic ground state of $\text{TbInO}_3$. This work establishes a measurement technique using superconducting resonators to probe thin films of frustrated magnets, and applies this technique towards building a coherent understanding of the magnetic properties of $\text{TbInO}_3$.

\end{abstract}

\maketitle


\section{Introduction}

Broken-symmetry ground states are associated with fascinating properties such as ferromagnetic order and superconductivity. Less-studied but equally interesting are states in which global symmetry breaking is prevented by frustration and quantum fluctuations. 
Quantum spin liquids (QSLs) are canonical realizations of this, where the lattice geometry imposes frustration among localized magnetic moments, and the exchange interaction between moments introduces quantum fluctuations \cite{balentsSpinLiquidsFrustrated2010a, broholm2020quantum}.

In theoretical models, the effective dimensionality of the spin lattice plays a key role in the realization of QSLs. In 1D antiferromagnetic spin chains, quantum fluctuations alone suffice to yield QSL states, as corroborated experimentally in a variety of materials \cite{lakeQuantumCriticalityUniversal2005, lecheminantOnedimensionalQuantumSpin2012}. In contrast, in 3D antiferromagnets (AFs), an ordered N\'{e}el ground state can be stabilized even in the presence of frustration and quantum fluctuations, in accordance with the Mermin-Wagner theorem \cite{lee2006doping, lakeQuantumCriticalityUniversal2005}. It is in 2D layered AFs with lattices such as triangular, honeycomb, and kagome that the tension between quantum fluctuations and magnetic ordering can lead to QSL ground states \cite{balentsSpinLiquidsFrustrated2010a, broholm2020quantum}. Hence, 2D AFs with frustrated lattice geometry appear promising targets for experimentally realizing QSLs. Accordingly, materials such as Herbertsmithite \cite{norman2016colloquium}, 2D-organic salts \cite{mikschGappedMagneticGround2021}, $\alpha$-RuCl$_3$ \cite{bruin2022robustness, li2021giant}, 1T-TaSe$_2$ \cite{ruanEvidenceQuantumSpin2021}, YbMgGaO$_4$ \cite{paddison2017continuous}, and KYbSe$_2$ \cite{scheieProximateSpinLiquid2024} have been extensively investigated with a wide range of techniques.

\begin{figure*}[t]
    \centering
    \includegraphics[width=0.9\textwidth]{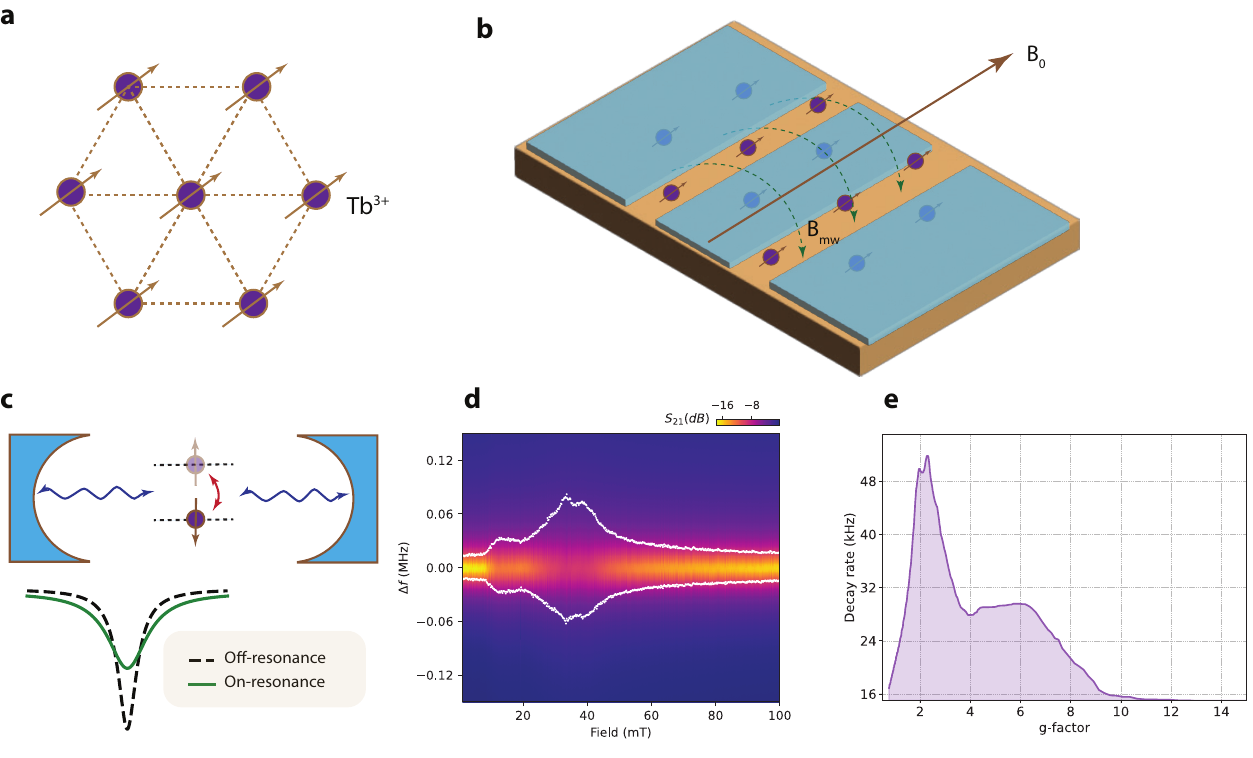}
    \caption{Device schematic and basic principles of spin resonance detection: (a) Tb$^{3+}$ ions with a partially filled $4f^8$ subshell form a frustrated triangular lattice. At low temperatures ($T \lesssim 10\,\si{\kelvin}$), each Tb$^{3+}$ is in a doublet  state. (b) Schematic of coplanar microwave resonator fabricated by etching a NbTiN superconductor (blue) deposited on a TbInO$_3$ film (golden). The microwave magnetic field $B_{mw}$ points along paths concentric with the center pin of the waveguide. An external static magnetic field $B_0$ is applied in the plane of the film such that the static field is perpendicular to the microwave field. (c) A spin embedded in a coplanar microwave resonator interacts with the resonator's electromagnetic field. When the frequency of the resonator is equal to the Zeeman splitting of a spin or the energy of a collective excitation, a resonant exchange occurs between photons in the resonator and the spins. This process causes an increase in the decay rate of the resonator. (d) $S_{21}$ (dB) of the microwave feedline coupled to the resonator as a function of $B_0$ and frequency. The white dots are a guide to the eye and track the resonator width. (e) The decay rate of the resonator as a function of an extracted $g$-factor, where $g = \frac{h f}{\mu_B B_0}$ where $f$ is the frequency of the resonator and $\mu_B$ is the Bohr magneton. Prominent broad resonances are centered around $g = 2.1$ and $g = 4 - 9$.}
    \label{fig1}
\end{figure*}

Here, we focus on a layered hexagonal antiferromagnet, TbInO$_3$, where the rare-earth Tb$^{3+}$ ion has a local magnetic moment. Magnetic susceptibility of bulk single-crystal TbInO$_3$ anomalously continues to follow a naive Curie-Weiss form down to 450 mK, far below the Curie-Weiss temperature~\cite{clarkTwodimensionalSpinLiquid2019a}. Inelastic neutron scattering measurements on both powder \cite{clarkTwodimensionalSpinLiquid2019a} and single-crystal \cite{kimSpinliquidlikeStatePure2019} samples have confirmed the lack of conventional magnetic order at temperatures down to two orders of magnitude lower than the energy scale set by the exchange interaction ($\theta_{\text{CW}} \sim 15\,K$). This failure to order has been attributed to the arrangement of Tb moments in a frustrated 2D lattice, and has been suggested to reflect a spin liquid ground state~\cite{kim2019spinliquid}. An additional subtlety is that Tb sites are spatially shifted in a periodic pattern along the c-axis of the unit cell, leading to improper ferroelectricity~\cite{kim2019spinliquid,nordlander2025signatures}. This combination of magnetic frustration, large spin-orbit coupling, lattice distortion, and the resultant local crystal fields is unique to TbInO$_3$ among spin liquid candidate materials.  

Whereas single crystals of TbInO$_3$ have been extensively studied since their first synthesis in 2018, growth of thin films could bring important opportunities such as using interfaces to introduce novel functionalities or break symmetries, and for inducing strain \cite{rameshCreatingEmergentPhenomena2019}. Furthermore, nanofabrication of functional devices is far easier on thin films compared to bulk crystals. Motivated by these advantages, epitaxial thin films of TbInO$_3$ were recently synthesized for the first time~\cite{nordlander2025signatures, talit2025structural}. Characterizing magnetic order and fluctuations in thin films requires new approaches, as workhorse measurements of heat capacity and neutron scattering are technically challenging to apply to thin films. Near-DC magnetic susceptibility measurements on such films are challenging but possible. Recently, both macroscopic and scanning SQUIDs were used to characterize TbInO$_3$ films and found that the Tb moments do not fully order down to 44\,\si{\milli\kelvin} \cite{nordlander2025signatures}.

Here, we set out to develop a complementary probe of magnetic order and fluctuations in thin films, and apply it to TbInO$_3$. Inspired by progress in the field of hybrid quantum systems based on circuit quantum electrodynamics \cite{clerk2020hybrid}, we develop a microwave magnetic resonance technique which is compatible with thin films and capable of measurements down to milliKelvin temperatures. Using planar superconducting resonators, we probe spin and collective magnetic excitations in TbInO$_3$ thin films. From a resonant response as a function of magnetic field, we extract two distinct g-factors associated with the moments in TbInO$_3$, and compare them with a simple model based on superpositions of crystal-field eigenstates of Tb$^{3+}$ moments. Why two distinct g-factors? TbInO$_3$ in both bulk and thin film form has been shown to be an improper ferroelectric \cite{kim2019spinliquid, nordlander2025signatures}. Since the g-factor is a sensitive probe of the local environment of a magnetic moment, we associate the two g-factors with the two types of Tb sites under the known ferroelectric distortion. We extract the integrated spectral weight of each resonant feature, which is proportional to the magnetic susceptibility of the associated population of moments. In this way we show that Tb moments fluctuate down to $20\,$mK, extending previous reports of magnetic frustration in this material system to almost three orders of magnitude below the exchange interaction energy scale ($\sim 11\,\si{\kelvin})$.

\begin{figure*}[t]
    \centering
    \includegraphics[width=0.9\textwidth]{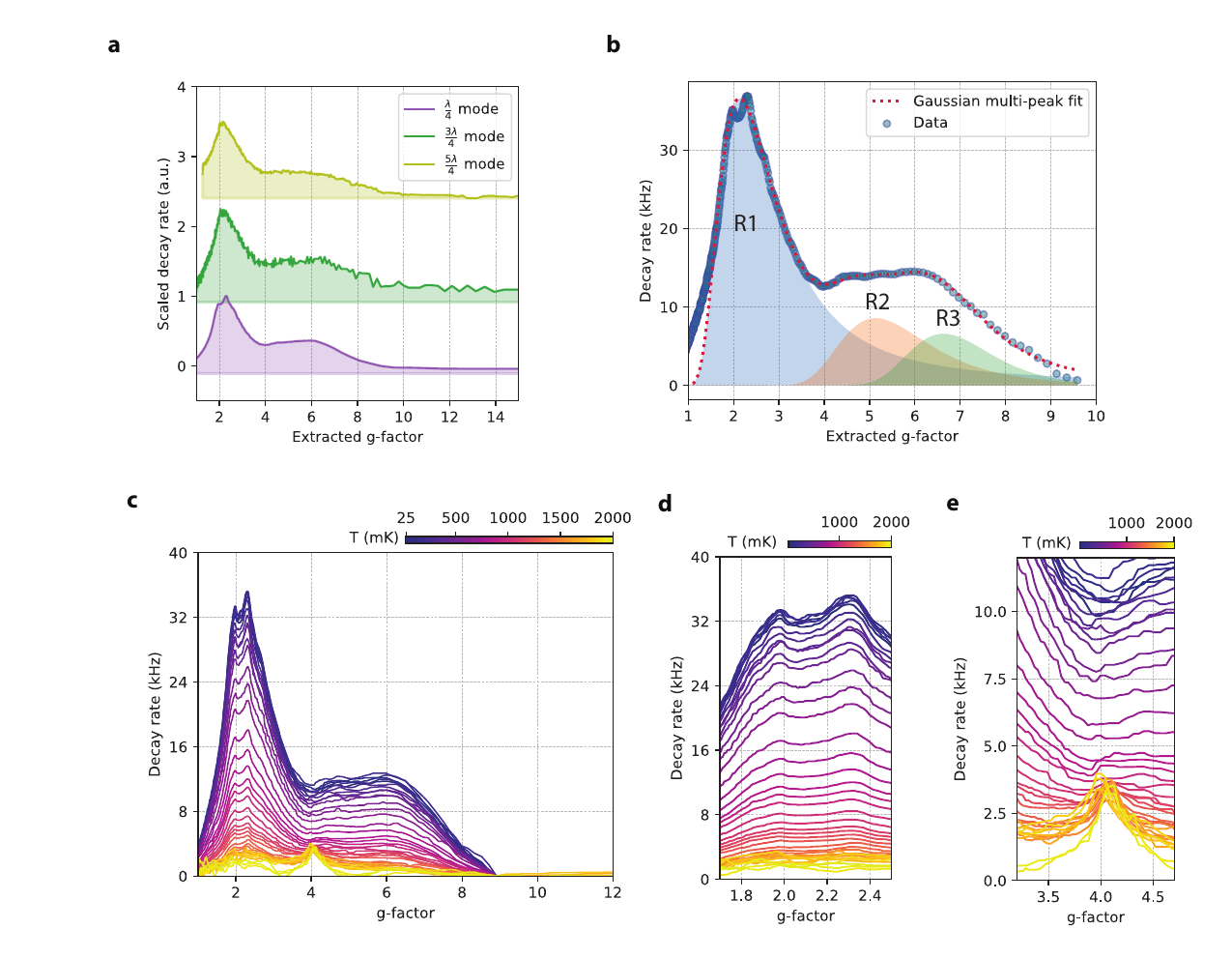}
    \caption{Frequency and Temperature dependence of the spin resonance signal: (a) Decay rate of the resonator measured as a function of magnetic field for $\lambda/4$, 3$\lambda/4$ and $5\lambda/4$ modes of the same resonator. The corresponding frequencies are $f$, $3f$ and $5f$. An extracted g-factor $\left( g = h f/(\mu_B B_0) \right)$ is used as the x-axis. The response is qualitatively similar across the three modes when plotted as a function of g. (b) A multipeak Gaussian model with three major components gives a good fit to the signal measured at the lowest temperature ($\sim 18\,\si{\milli\kelvin}$). We denote these three major components as R1 ($g = 2.13 \pm 0.01$), R2 ($g = 5.1 \pm 0.9$) and R3 ($g = 6.6 \pm 0.5$). (c) Temperature dependence of the decay rate vs g-factor as a function of sample temperature from $18\,\si{\milli\kelvin}$ to $2\,\si{\kelvin}$. (d) Two narrow resonances at $g = 1.95$ and $g = 2.3$ are observed on top of R1.  (e) A third narrow resonance at $g=4.1$ has a very different phenomenology: suppressed for $T \lesssim 800\,\si{\milli\kelvin}$, but present at higher temperatures.}
    \label{fig2}
\end{figure*}

\begin{figure*}[t]
    \centering
    \includegraphics[width=0.9\textwidth]{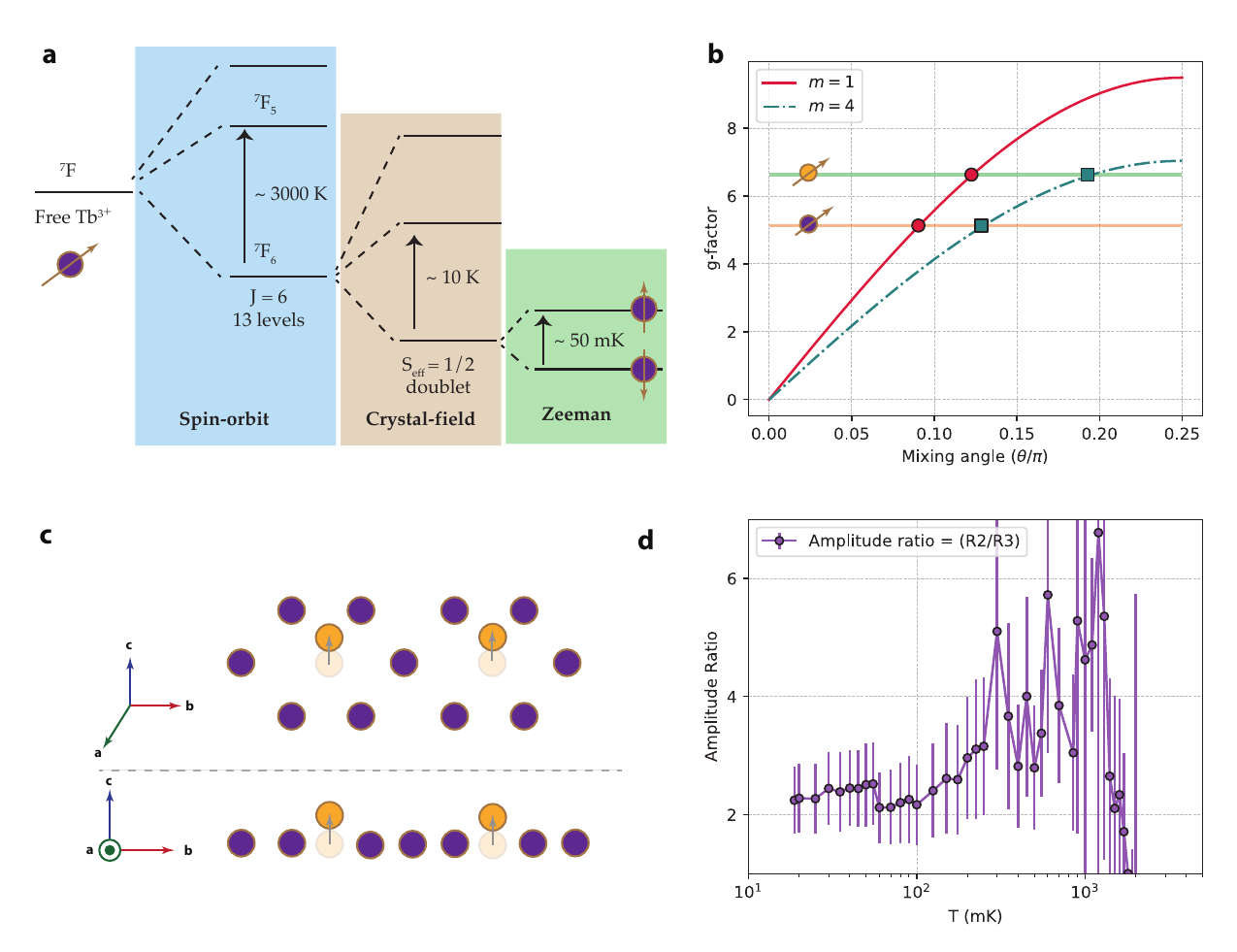}
    \caption{Crystal-field model and c-axis displacement of Tb$^{3+}$ ions (a) The hierarchy of energy scales that govern the ground state of a single Tb$^{3+}$ ($4f^8,\, L = 3, \, S=3$) ion, neglecting coupling to other Tb$^{3+}$ sites. A large spin-orbit coupling from the high-Z Tb nucleus implies $J = L+S$ is a good quantum number and the atomic term $^7F_6$ with $J = 6$ is the ground state. $J=0,...,5$ states are separated by $\sim 3000\,\si{\kelvin}$ from the ground state. The $^7F_6$ state with $J=6$ is a collection of $2J+1 = 13$ levels. The crystal field from nearby ions further splits these into a doublet ground state separated from the lowest-lying crystal field excited level at $10\,\si{\kelvin}$. We apply an external magnetic field to finally split this doublet, adjusting the field so the splitting matches the resonator frequency. Note that the splittings are not drawn to scale.(b) $g$-factor calculated using a simple crystal-field model that mixes the $m_J=-6,...,6$ components of the $J=6$ moment, yielding a ground-state doublet built from either $m_J = \pm 1,\pm 2$ states (solid line) or $m_J = \pm 4,\pm 5$ states (dash-dot line). The mixing angle $\theta$ parametrizes the superposition between the two constituent $m_J$ states in each of these scenarios. For either assignment of the constituents of the ground-state doublet, the two $g$-factors for R2 and R3 as extracted in the previous figure can be matched by setting two different values of the mixing angle, as represented by red circles ($m_J = \pm 1,\pm 2$) and teal squares ($m_J = \pm 4,\pm 5$). (c) Displacement of $1/3$ Tb ions along the c-axis leads to a distortion of the triangular lattice and hence the local crystal field environment. As a result, two flavors of Tb ions exist, each with a distinct crystal-field environment. (d) Amplitude ratio of the two Tb-moment peaks R2 ($g = 5.1 \pm 0.9$) and R3 ($g = 6.6 \pm 0.5$), extracted from the Gaussian fit model. At low temperatures, the amplitude ratio is close to 2. This suggests two classes of Tb moments with distinct g-factors are present in the film in a 2:1 ratio of concentration.}
    \label{fig3}
\end{figure*}

\section{Probing magnetism in \boldmath\texorpdfstring{T\MakeLowercase{b}I\MakeLowercase{n}O$_3$}{TbInO3} by microwave spin resonance}
 We grow TbInO$_3$ films on a yttria stabilized zirconia (YSZ) $\langle 111 \rangle $ substrate using molecular beam epitaxy (MBE) as described in \cite{nordlander2025signatures}. The YSZ substrate ensures epitaxial growth of TbInO$_3$ with the c-axis perpendicular to the substrate surface. We then use reactive magnetron sputtering with Nb and Ti targets in a Ar/N$_2$ environment to deposit superconducting NbTiN films onto the TbInO$_3$ films. Finally, we use photolithography followed by reactive plasma etching to pattern the superconducting film into a microwave feedline capacitively coupled to a $\lambda/4$ co-planar waveguide (CPW) resonator. Additional details on crystal structure, growth, fabrication and measurement are provided in the methods section and supplementary information \cite{supp}.

TbInO$_3$ has a hexagonal unit cell in which layers of magnetic Tb$^{3+}$ ions are separated by layers of non-magnetic [InO$_5$]$^{7-}$ corner-sharing trigonal biprism polyhedra \cite{clarkTwodimensionalSpinLiquid2019a, kimSpinliquidlikeStatePure2019, yeCrystalfieldExcitationsVibronic2021b}. A c-axis lattice constant of $12.3\,\si{\angstrom}$, far larger than Tb-Tb intra-layer separation $ 3.65\,\si{\angstrom}$, coupled with the compactness of f-orbitals, implies quasi-2D interactions between adjacent Tb moments. Fig.~\ref{fig1}a schematically shows an ideal triangular lattice of Tb ions. Each Tb is in a $3+$ oxidation state, corresponding to an incompletely filled $f$-subshell with electronic configuration $4f^{8}$. For bulk crystals, a careful analysis of powder and single-crystal magnetic susceptibility along with assignment of crystal fields based on neutron scattering and Raman spectroscopy have led to a common conclusion that an effective non-Kramers magnetic doublet state exists at low temperatures ($\lesssim 5\,\si{\kelvin}$) \cite{clarkTwodimensionalSpinLiquid2019a, kimSpinliquidlikeStatePure2019, yeCrystalfieldExcitationsVibronic2021b}. The failure of this doublet to globally order has been attributed to frustration associated with antiferromagnetic coupling on a triangular lattice. We probe this doublet state using microwave spin resonance. 

Within the mode volume of a microwave resonator, electric and magnetic fields at the resonator frequency are strongly enhanced. These fields can couple to spins and other collective excitations of a material placed within the mode volume, so the field enhancement can compensate for the low volume of the material to be probed. Such a resonator can be described as a damped harmonic oscillator with frequency $f$ and energy loss rate $\kappa$. The field enhancement is given by the quality factor $Q= \frac{f}{\kappa}$. Resonators in planar rather than three dimensional geometry are particularly well suited to coupling to thin films: the electromagnetic field produced by the resonator falls off on a length scale set by its lithographic features, typically microns, so a resonator fabricated or placed atop a thin film of interest will probe that film as well as the top portion of the underlying substrate, but will be insensitive to the bulk of the substrate. We deposit NbTiN on TbInO$_3$ films, then lithographically pattern the NbTiN to form superconducting planar resonators. Fig.~\ref{fig1}b schematically shows a coplanar waveguide with a center conductor surrounded by a ground plane on top of the TbInO$_3$ film. Since the $22\,\si{\nano\meter}$ thickness of the film is much smaller than the $10\,\si{\micro\meter}$ width of the center conductor, Tb spins through the entire film thickness are coupled to the resonator, as are any free spins at interfaces or within the top $\approx 10\,\si{\micro\meter}$ of the substrate. We achieve quality factors $Q = \frac{f}{\kappa} \sim 10^5$, comparable to those reported for microfabricated planar superconducting resonators on more conventional substrates~\cite{bienfait2016reaching}. 

A spin interacting with an external magnetic field is described by the Zeeman Hamiltonian:
$H = g \mu_B \vec{B} \cdot \vec{S} $
where $g$ is the g-factor, $\mu_B$ is the Bohr magneton, $\vec{B}$ is an externally applied magnetic field and $\vec{S}$ is the quantum mechanical spin operator. We apply an external static magnetic field $B_0$ in the plane of the film. As a result, the two spin eigenstates are split by $\Delta E = g \mu_B B_0$. We typically apply fields on the order of $10-300\,\si{\milli\tesla}$, corresponding to splittings $\frac{\Delta E}{h} = 0.28-8.4\,\si{\giga\hertz}$ (for $g=2$), making resonators designed in the microwave $\si{\giga\hertz}$ range ideal for our measurements. We use a resonator with a fundamental frequency $f = 1.064\,\si{\giga\hertz}$. Upon tuning the external magnetic field, spins with different g-factors are independently coupled to the same resonator with the resonance condition $h f = g \mu_B B_0$. In resonance, the microwave field in the resonance drives transitions between the two spin eigenstates, which have a nonzero population difference set by temperature as $\tanh\left(\frac{h f}{2 k T} \right)$. As a result of these transitions, the decay rate $\kappa$ of the resonator gets an additional contribution which can be measured (see Fig.~\ref{fig1}c). The resultant spin-photon hybridization also causes a shift in the resonator frequency, but since we are in the weak coupling limit \cite{clerk2020hybrid} this frequency change is too small to be detected. Hence, we rely on measuring the decay rate of the resonator as a function of $B_0$ to detect spin resonance. 

Fig.\,\ref{fig1}d shows the raw transmission coefficient $S_{21} (dB)$ of our microwave setup as a function of frequency and magnetic field. At a given magnetic field, a narrow dip appears at the resonator frequency $f_0$. In the plot, we shift the center of the resonance dip to zero at each magnetic field to highlight changes in the resonator width. The resonator width quantified into a decay rate of the resonator (see Methods for fitting and analysis of the resonator profile) changes as a function of the magnetic field. When the magnetic field is near a resonance condition for spins in the mode volume ($h f = g \mu_B B$), the resonator decay rate is enhanced. To make a transparent connection to the underlying spins interacting with the resonator, we plot the decay rate as a function of an extracted g-factor defined as $g=\frac{h f}{\mu_B B_0}$ in fig.~\ref{fig1}e. We observe a prominent broad peak centered around $g = 2.1$, and another broad peak with a spread of g-factors from $4-9$.

A $\lambda/4$ CPW resonator also has higher modes at $3f$ and $5f$, in our case at $3.2\,\si{\giga\hertz}$ and $5.3\,\si{\giga\hertz}$, respectively. In Fig.~\ref{fig2}a, we plot the decay rate for $f$, $3f$ and $5f$ modes as a function of the extracted g-factor. We observe that resonances (enhanced decay rate) occur at the same extracted g-factor for the three resonator modes: a feature that occurs at magnetic field $B_0$ for a resonator mode at frequency $f$ occurs at magnetic fields $3 B_0$ and $5 B_0$ when measured at $3f$ and $5f$. This pattern allows us to rule out explanations other than magnetic resonance, for example loss due to superconducting vortices or parasitic cavity modes. Hence, a linear in magnetic field splitting described by a constant g-factor is indeed a good phenomenological descriptor of the observed resonances.

The width of a spin resonance signal is tied to the spin relaxation time. A narrow resonance ($< 1\,\si{\milli\tesla})$ typically indicates localized spins on defect centers with long relaxation times, whereas broader resonances indicate that other interactions such as dipole-dipole and anisotropic exchange determine the linewidth. Moreover, the broadening mechanism also determines the functional form of the lineshape. For homogeneously-broadened lines, a Lorentzian lineshape is commonly used. On the other hand, when the local environment of individual sites is different, an inhomogeneously broadened line with a Gaussian lineshape is used.  In Fig.~\ref{fig2}b, a multi peak Gaussian with three components gives the best fit to the data. Fitting with two Gaussians is not sufficient to capture the broad shoulder from g=4-9 and with three peaks, we see a good agreement between the data and the fit. We denote these three major components as R1 ($g = 2.13 \pm 0.01$), R2 ($g = 5.1 \pm 0.9$) and R3 ($g = 6.6 \pm 0.5$). The R1 resonance is also observed on a second resonator fabricated on the same chip, with the TbInO$_3$ film removed by ion milling to extract the YSZ substrate response \cite{supp}. Hence, R1 must be a result of defects or impurities in the 
YSZ substrate. On the other hand, g-factors corresponding to R2 and R3 are not seen from the resonator coupled only to the YSZ substrate, so they must arise from the TbInO$_3$ film or in principle its interface with the substrate.

Fig.~\ref{fig2}c shows the temperature dependence of the spin resonance signal from $18\,\si{\milli\kelvin}$ to $2\,\si{\kelvin}$. With increasing temperature, the response is diminished by the decrease in polarization of the participating spins in the small applied magnetic field. We first focus on the identification of the g-factors corresponding to the narrow lines in our dataset. We show these narrow lines originate from impurities or defects in the superconducting film and the YSZ substrate. Fig.\ref{fig2}d and e show a zoom-in around the g-factor ranges where narrow lines are observed. The $g = 1.9$ line (Fig.\ref{fig2}d) also occurs in a separate NbTiN resonator fabricated on silicon, showing that it does not reflect spins in either our thin film of interest or the YSZ substrate. We associate this feature with Nb$^{4+}$ defects present on the surface of the superconducting film \cite{supp}. These defect spins have been recently reported in NbTiN/sapphire films through a similar resonator measurement \cite{bahr2024improving}. The $g=2.3$ line (Fig.\ref{fig2}d) also occurs in the second resonator on the same YSZ chip, where the TbInO$_3$ film was removed by ion milling to aid in background subtraction. However, it does not occur in another resonator on an otherwise-unprocessed YSZ substrate (no TbInO$_3$ ever deposited) \cite{supp}. This suggests the feature arises from modification of the YSZ substrate during the growth of the TbInO$_3$. YSZ has been shown to conduct oxygen ions at high temperatures \cite{zakaria2020review}. A g-factor slightly larger than 2 is consistent with an oxygen vacancy defect with spin-orbit coupling of intermediate strength in comparison to crystal fields \cite{slichter2013principles, costantini2011paramagnetic}. Finally, a third sharp resonance at $g=4.1$ has a temperature dependence quite different from that of all our other resonances: instead of fading with increasing temperature, it is absent for temperatures lower than $\sim 800\,\si{\milli\kelvin}$ and emerges only at higher temperatures. Since ESR spectroscopy of Fe impurities in other materials has identified $g=4.1-4.3$ resonances as associated with the Fe$^{3+}$ state \cite{castner1960note, vasyukovManifestationNoncentralityEffect2011}, we identify this resonance with Fe impurities in the YSZ substrate. From the magnitude of the signal at 2\,\si{\kelvin}, we estimate a 60-840 ppm concentration of Fe impurities in the YSZ substrate \cite{supp}. 
We speculate that the disappearance of the resonance signal below $800\,\si{\milli\kelvin}$ could be caused by the Fe$^{3+}$ impurity moments entering a spin glass state. 

\begin{figure}
    \centering
    \includegraphics[width=0.9\linewidth]{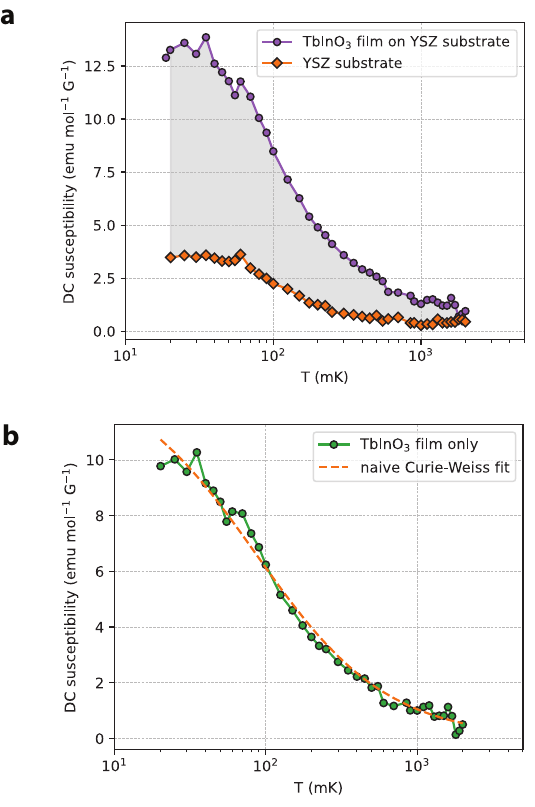}
    \caption{Magnetic susceptibility from $20\,\si{\milli\kelvin}-2\,\si{\kelvin}$ (a) Magnetic susceptibility inferred by calculating the area under the decay rate vs magnetic field curve. Since the resonator is also sensitive to signals from the substrate, we fabricate a second resonator on the same chip in a region where the TbInO$_3$ has been removed by ion milling. The susceptibility is reported on the same y-axis scale with TbInO$_3$ plus substrate (purple squares) and substrate alone (orange diamonds) at temperatures from $20\,\si{\milli\kelvin}-2\,\si{\kelvin}$. (b) The difference between the two values at each temperature is our best measure of the susceptibility of the TbInO$_3$ film. We observe a non-saturating susceptibility which is well-described by a naive Curie-Weiss law down to $50\,\si{\milli\kelvin}$, below the Curie-Weiss scale from that same fit, and 220 times below the Curie-Weiss scale extracted from higher-temperature data. Susceptibility may start to saturate between $20$ and $50\,\si{\milli\kelvin}$.}
    \label{fig4}
\end{figure}

\section{Model for ground state eigenstates}
Having assigned the narrow lines ($g\approx 1.9, 2.3,$ and 4.1) to various interfacial defects and bulk impurities, we now focus on the origin of broad resonances (R2 and R3) in our spectra. We start with the energy level scheme of each constituent Tb ion. Rare earth ions are simpler to model than transition metal ions, as they have a clear hierarchy in the magnitude of energy scales (Coulomb $\gg$ spin-orbit $\gg$ crystal-fields $\gg$ hyperfine). This hierarchy makes the identification of a ground state manifold unambiguous from first principles.

Fig.~\ref{fig3}a shows the hierarchy of energy scales from the atomic levels to the doublet ground state for each Tb$^{3+}$. The $3+$ oxidation state has an outer electronic configuration of $4f^8$. By Hund's rules, we get the total orbital angular momentum $L = 3$ and total spin angular momentum $S=3$. The high atomic number of the rare earth atoms leads to a large spin-orbit interaction which couples L and S into a total angular momentum $J = L-S,...,|L+S|$. Since the f-subshell is more than half-filled, the sign of the spin-orbit interaction dictates that $J=6$ is the ground state. The $^7F_6 (^{2S+1}L_J)$ ground state is separated from $J=5$ state by $\sim 3000\,\si{\kelvin}$ \cite{abragam2012electron}. The $J=6$ level itself is compromised of $2J+1 = 13$ $m_J$ states with $m_J = -6,...,6$. Crystal fields imparted by the local environment of the ion cause a splitting of the $J=6$ multiplet into singlet and doublet states \cite{yeCrystalfieldExcitationsVibronic2021b}. Inelastic neutron scattering and Raman spectroscopy on TbInO$_3$ single crystals have established a doublet ground state separated by a low-lying crystal field (CF) level at roughly $10\,\si{\kelvin}$ \cite{clarkTwodimensionalSpinLiquid2019a, kimSpinliquidlikeStatePure2019, yeCrystalfieldExcitationsVibronic2021b}. Since our experiment is conducted in the $20\,\si{\milli\kelvin}-2\,\si{\kelvin}$ range, the higher CF levels are unoccupied and the physics is dominated by the properties of the ground state doublet. When we apply a static magnetic field, this doublet state is split, and the field can be tuned so this splitting matches the resonator frequency $f = 1\,\si{\giga\hertz} \sim 50\,\si{\milli\kelvin}$.

Which $m_J$ states participate in this doublet, and their relative weight, remain to be identified. The trigonal symmetry of each Tb site restricts the $m_J$ quantum numbers that can participate to give a doublet state \cite{yeCrystalfieldExcitationsVibronic2021b}: $m_J = 0,\pm 3, \pm 6$ states are required to be singlets. We are left with $m_J = \pm 1, \pm 2, \pm 4, \pm 5$, an eigenspace spanned by 8 states, which can potentially exist in a superposition to give a doublet state. At this point, the g-factors extracted from our spin resonance measurements provide further constraints. Spin resonance involves absorption of photons which carry angular momentum $\pm 1 \hbar$. This leads to another constraint that doublets observed in ESR must incorporate $m_J$ states which differ by $\pm1$. 

We propose an ansatz that mixes the $m_J$ states, parameterized by a single mixing angle $\theta$: 
\begin{align}\label{eqn-ans}
     |\psi_{m,+}(\theta)\rangle &= c |\phi_{m,+}\rangle +s|\phi_{m+1,+}\rangle \\
    |\psi_{m,-}(\theta)\rangle &=  c |\phi_{m,-}\rangle  - s|\phi_{m+1,-}\rangle 
\end{align}
where $|\psi_{m,\pm}\rangle$ is a candidate doublet ground state, $c=\cos(\theta)$, $s=\sin(\theta)$ and $|\phi_{m,\pm}\rangle = \frac{1}{\sqrt{2}} \left(|m\rangle \pm |-m\rangle \right)$. Each $|\psi_{m,\pm}\rangle$ state mixes $\pm m$ and $\pm (m+1)$ states: $|\psi_{1,\pm}\rangle$ mixes $m=1$ and $m=2$, whereas  and $|\psi_{4,\pm}\rangle$ mixes $m=4$ and $m=5$. Fig.~\ref{fig3}b shows the calculated g-factor as a function of mixing angle for $m=1$ and $m=4$.

Recall that we observe two different g-factors for the resonances R2 and R3 (Fig.~\ref{fig2}b.) The g-factor is a sensitive probe of the local environment of each Tb site, so the different g-factors may reflect two non-equivalent Tb sites in the film. TbInO$_3$ belongs to a class of hexagonal ABO$_3$ compounds which includes hexagonal manganites \cite{vanakenOriginFerroelectricityMagnetoelectric2004} and ferrites \cite{dasBulkMagnetoelectricityHexagonal2014a} that undergo a structural transition into a improper ferroelectric phase. The existence of such a structural distortion and the corresponding domain structure in TbInO$_3$ have been established by scanning transmission electron microscopy (STEM) imaging of bulk single crystals and thin films\cite{nordlander2025signatures,el2022multi}. Fig.~\ref{fig3}c shows a schematic of the Tb site distortion: $1/3$ of the Tb sites are shifted along the c-axis. If we posit that the different g-factors are associated with different mixing angles on the inequivalent sites, R2 and R3 both lie in the range that can be accounted for by either $|\psi_{1,\pm}\rangle$ or $|\psi_{4,\pm}\rangle$, for suitable choices of mixing angles in each case: in Fig.~\ref{fig3}b, matching mixing angles are marked by circles for $|\psi_{1,\pm}\rangle$, and squares for $|\psi_{4,\pm}\rangle$. 

Fig.~\ref{fig3}d shows the amplitude ratio of R2 and R3 as extracted from multi-peak Gaussian fitting. For a given temperature, amplitude is proportional to the number density of the moments participating in the resonance. We experimentally observe an amplitude ratio close to 2. This matches a naive expectation of the density ratio of two classes of Tb sites that are rendered inequivalent by the improper ferroelectric distortion. The fact that the sample does not consist of a single ferroelectric domain initially gave us pause: STEM imaging has shown two kinds of structural domains with a typical domain size ranging from 20-40\,\si{\nano\meter} \cite{nordlander2025signatures}. Since the resonator is sensitive to a all spins within a characteristic length set by the resonator dimensions ($\sim 5-10\,\si{\micro\meter}$), we expect our measurement to represent an average over these structural domains. The structural domains are related to each other by a reflection in the $ab$ plane, which transforms $m_J$ quantum numbers as $m \rightarrow -m$. Since our ansatz for the ground state doublets involves symmetric and anti-symmetric combinations of $\pm m$ states, the wavefunctions $|\psi_{m,\pm}\rangle$ of the doublet states only acquire a global phase under this transformation. This means that not only the g-factors of the two inequivalent Tb sites but also the relative density of sites associated with those g-factors should be preserved across domain boundaries, consistent with our observation of an amplitude ratio close to 2 rather than 1.
 
We have aimed to explain the experimental observations with a minimum set of parameters, so our ansatz is designed to capture only the simplest superpositions consistent with visible transitions in ESR. The actual doublets could combine not just two but three or even all four of the allowed magnitudes of magnetic quantum numbers $|m|=1, 2, 4, 5$. Experiments which can measure the complete angular dependence of the g-factor or even g-factors along high-symmetry directions of the crystal lattice would allow more specific assignment of the doublet ground states.

\section{Extraction of Magnetic Susceptibility}
The integrated area under the spin resonance line as a function of magnetic field is directly proportional to the low-frequency magnetic susceptibility \cite{polash2023electron}. By applying an in-plane magnetic field, we measure the in-plane (ab plane) magnetic susceptibility as a function of temperature. Note that recently-reported measurements on TbInO$_3$ films using a scanning SQUID were sensitive to the out-of-plane (c-axis) susceptibility down to 44\,\si{\milli\kelvin}, whereas measurements of in-plane susceptibility using a commercial SQUID were performed down to 400\,\si{\milli\kelvin} \cite{nordlander2025signatures}. Here, we extend the temperature range for in-plane susceptibility down to 20\,\si{\milli\kelvin}. As the YSZ substrate itself has magnetic impurities and defects, we ion-mill off a section of the TbInO$_3$ film and fabricate a second resonator device on the same chip, with identical geometrical parameters. Such a device on the same chip allows us to subtract the background susceptibility of the YSZ substrate and thus extract a susceptibility of the TbInO$_3$ film alone. We convert the spectral weight from the resonance curve into emu mol$^{-1}$ G$^{-1}$ units for direct comparisons to DC susceptibility reported by others \cite{supp}. In temperature ranges overlapping other measurements, our extracted in-plane susceptibility is within a factor of 2-3 of previously reported values \cite{clarkTwodimensionalSpinLiquid2019a, nordlander2025signatures}. We attribute any remaining inconsistency in the absolute magnitude of the susceptibility to approximations made by us in estimating the participation ratio of the spins coupled to the resonator. 
 
Fig.~\ref{fig4}a shows the extracted susceptibility for the TbInO$_3$/YSZ device and the control YSZ device. For both cases, we observe susceptibility increases with decreasing temperature. Above 2K, the spin resonance signal is difficult to detect (see Fig.~\ref{fig2}c) due to decrease in thermal polarization of the participating spins and the reduction in quality factor of the resonator at higher temperatures. For all temperatures between $20\,\si{\milli\kelvin}$ and 2K, the extracted susceptibility from the TbInO$_3$/YSZ device is strictly larger than that of the YSZ device. Fig.~\ref{fig4}b shows the result of subtraction between these two curves as the susceptibility we attribute to the TbInO$_3$ film alone. The susceptibility increases with decreasing temperature at least down to 50 mK, with possible indication of saturation below that, and can be fit to a naive Curie-Weiss form $\left(\chi(T) = \frac{C}{T - \theta^{LT}_{\text{CW}}}\right)$ as shown in Fig.~\ref{fig4}b. From the fit to data between 20\,\si{\milli\kelvin} and 2\,\si{\kelvin}, we obtain a Curie-Weiss constant $\theta^{LT}_{\text{CW}} = -86 \pm 5\,$ mK, far lower in magnitude than that extracted from higher temperature (2\,\si{\kelvin}-300\,\si{\kelvin}) measurements, $\theta_{\text{CW}} \sim -11\,\si{\kelvin})$. A similar reduction in $|\theta_{\text{CW}}|$ was reported for polycrystalline TbInO$_3$ when the researchers fit to their lowest-temperature data from 400\,\si{\milli\kelvin} to 1\,\si{\kelvin} rather than to their higher-temperature data \cite{clarkTwodimensionalSpinLiquid2019a}. 

When an antiferromagnet orders, susceptibility saturates or decreases as temperature is lowered below the ordering temperature. Between $50\,\si{\milli\kelvin}$ and our base temperature of $20\,\si{\milli\kelvin}$, susceptibility may be starting to saturate, or the sample may not fully thermalize during ESR measurements. However, down to $50\,\si{\milli\kelvin}$ we don't observe any decay or saturation of the susceptibility, so we conclude that no ordering transition occurs in TbInO$_3$ down to that temperature, 220 times lower than the Curie-Weiss scale $|\theta_{\text{CW}}| \sim 11\,\si{\kelvin})$ extracted from high-temperature ($2-300\,\si{\kelvin}$) susceptibility measurements \cite{supp, nordlander2025signatures}. The degree of frustration in magnetic materials is commonly quantified by a frustration index $f = |\theta_{\text{CW}}|/T_{\text{ord}}$, where $T_{ord}$ is the ordering temperature \cite{Ramirez1994, supp}. Our measurements show that the frustration index in TbInO$_3$ is at least 220, one of the highest values reported in the search for candidate materials to host QSL states (see \cite{supp} for a table of frustration indices). Our microwave measurements of in-plane susceptibility complement recent commercial and scanning SQUID measurements of these TbInO$_3$ films \cite{nordlander2025signatures,supp}, not only probing susceptibility in a completely independent way but extending in-plane susceptibility an order of magnitude lower in temperature to match the range of out-of-plane susceptibility measurements. The scanning SQUID-based out-of-plane susceptibility peaks around $1\,\si{\kelvin}$ and slowly decreases from the peak value with further decreasing temperature \cite{nordlander2025signatures}. An exponential decrease in susceptibility along the moment axis that would be expected below an ordering transition is not observed. Coupled with our in-plane susceptibility results, this suggests that the Tb moments have an easy-plane anisotropy, consistent with previous studies on powder and single-crystal samples \cite{clarkTwodimensionalSpinLiquid2019a}.  

\section{Discussion and Outlook}
Transition metal and rare-earth oxides have historically played a prominent role in the discovery of correlated phases of matter. The presence of energetically favorable couplings between charge, spin, orbital and lattice degrees of freedom in such materials catalyzes novel ground states \cite{rameshCreatingEmergentPhenomena2019}. TbInO$_3$ is exemplary in this regard, as all these factors are at play in the determination of the ground state.

Fluctuating Tb moments with a high frustration index as shown by these measurements and others \cite{nordlander2025signatures, clarkTwodimensionalSpinLiquid2019a, kimSpinliquidlikeStatePure2019, yeCrystalfieldExcitationsVibronic2021b} indicate a high degree of magnetic frustration in TbInO$_3$. The geometric distortion of Tb sites differentiates two subsets of Tb ions. One subset occupies a honeycomb lattice, the other a triangular lattice, each of these being a lattice geometry prone to magnetic frustration. What the ultimate ground state of such a dual-flavored magnetic system on distinct frustrated lattices would look like remains an open question. Moreover, since the ground state eigenstates are composed of f-orbital states, the exchange interaction is likely strongly anisotropic i.e. bond direction dependent. A honeycomb lattice of Tb moments along with anisotropic exchange raises an interesting connection to the Kitaev honeycomb spin model, which displays canonical QSL states \cite{takagi2019concept}. TbInO$_3$ and other rare-earth hexagonal ABO$_3$ oxides with magnetic frustration could fruitfully join the more widely-studied iridates and $\alpha$-RuCl$_3$ \cite{trebst2022kitaev, clarkTwodimensionalSpinLiquid2019a} in this context.

Finally, we emphasize that the microwave resonator technique described here is widely applicable to a large class of insulating materials with magnetic frustration \cite{gardner2010magnetic}. Advances in the field of circuit quantum electrodynamics as implemented by planar resonators have enabled hybrid spin-photon magnetic systems \cite{clerk2020hybrid}. Such resonators can probe not only local, nearly-free spins but also collective deconfined excitations such as spinons that have been proposed in certain QSL models \cite{luo2018spinon, balents2020collective, povarov2022electron}. In conclusion, the combination of new measurements techniques and thin film growth provides a rich landscape for the discovery and characterization of frustrated magnets, including spin liquid candidates.    

\bibliography{refs, si-refs}

\begin{acknowledgments}
We thank Austin Kaczmarek, Aaron Sharpe, Mihir Pendharkar and Marc Kastner for their insightful comments. Experimental measurements, analysis, and film growth were supported by the Air Force Office of Scientific Research (AFOSR) Multidisciplinary Research Program of the University Research Initiative (MURI) under grant number FA9550-21-1-0429. Fabrication, including patterning of resonators, was performed at nano@stanford, supported by the National Science Foundation under Award ECCS-2026822. 
Development of the microwave techniques was supported by the Department of Energy, Office of Science, Basic Energy Sciences, Materials Sciences and Engineering Division, under Contract DE-AC02-76SF00515. JN acknowledges support from the Swiss National Science Foundation under Project No. P2EZP2-195686.
SSK acknowledges graduate financial support from the Knight-Hennessy fellowship.
\end{acknowledgments}

\section*{Methods}

\subsection*{Growth and Device Lithography}
Thin films of TbInO$_3$ were grown by MBE on YSZ substrates \cite{nordlander2025signatures}. NbTiN films were deposited by magnetron co-sputtering of Nb and Ti targets in a Ar/N$_2$ environment. Direct-write photo-lithography with $385\,\si{\nano\meter}$ light was used to expose AZ1512 resist ($1.4\,\si{\micro\meter}$) thickness and subsequently developed in MF319 developer solution for 1 min. After optical inspection and a $30"$ oxygen descum, the NbTiN film was etched in a reactive-ion etcher with a SF$_6$/Ar plasma. The resist was finally removed in acetone and isopropanol.

\subsection*{Cryogenic measurement setup}
Each chip with patterned resonators was glued on to a custom-designed copper sample holder with mini-SMP connectors for microwave excitation and measurement. The sample holder is screwed on to a microwave click-in plate at the bottom of a top-loading probe used with a commercial He3-He4 dilution cryostat (Leiden Cryogenics, CF-CCS81). The click-in plate is thermally anchored to the mixing chamber of the cryostat with a minimum base temperature of $14\,\si{\milli\kelvin}$. In the main text, we report the temperature of a calibrated ruthenium oxide thermometer in proximity to the sample as the temperature of the sample. The minimum temperature as measured by the sample thermometer is $18\,\si{\milli\kelvin}$. A $100\,\si{\ohm}$ resistive heater on the mixing chamber of the cryostat is used to control sample temperature from base to $2\,\si{\kelvin}$. A PID loop ensures temperature stability with an error less than $0.1\,\%$. Temperature stability was found to be crucial when sweeping the magnetic field at a given temperature as any change in temperature also leads to a change in the decay rate of the resonator by generation of thermal quasi-particles in the superconductor. 

Microwaves are sent to the sample holder via lossy CuNi semi-rigid coax cables thermally anchored at 4K, Still, 50mK and mixing chamber plates of the cryostat. The input line is heavily attenuated by cryogenic attenuators (XMA Corp.) with attenuation of 20 dB (4K), 20 dB (Still), 20 dB (50 mK) and 6 dB (mixing chamber) to thermalize the signal line of the coax with the shield. On the output side, a cryogenic isolator (Quinstar) is mounted at the mixing chamber and is further connected to a cryogenic microwave amplifier (Cosmic Microwave - CITLF3) with a gain of 35 dB in frequency range $1-5\,\si{\giga\hertz}$. A chain of three room temperature low-noise microwave amplifiers (Mini Circuits, ZX60-83LN-S+) provides additional gain of about 60 dB. Measurements are performed with a vector network analyzer (Rohde and Schwarz, ZNB8) with an IF bandwidth = 500 Hz and with 1-5 averages per frequency trace. A doubly-isolated superconducting magnet (American Magnetics) capable of supplying fields up to 10T mounted on the still plate ($\sim 1\,\si{\kelvin}$) is used to apply the static magnetic field. The field was ramped slowly $0.25-0.5\,\si{\milli\tesla\per\second}$ with a 1-3 second wait between adjacent field points to prevent any inductive heating and avoid sudden introduction of vortices in the superconducting film. OFHC Cu and brass screws were used in the construction of the sample holder to avoid any stray fields from magnetic components near the sample. This is crucial in reporting g-factors as we use the absolute value of the magnetic field along with the resonator frequency to calculate the g-factor.

\subsection*{Fitting of resonator $S_{21}$ and extraction of resonator properties}

The superconducting resonator is capacitively coupled in a hanger geometry to an on-chip broadband microwave feedline. The coupling strength to the feedline is quantified by a coupling quality factor $Q_c$ and it is chosen to be such that the resonator is critically coupled to the feedline i.e. internal quality factor $Q_i$ is such that $Q_i \sim Q_c$. This enables us to extract small changes in the width i.e. the decay rate of the resonator with high sensitivity.

The $S_{21}$ magnitude and phase as measured by the VNA is fit to a model for hanger resonators \cite{khalil2012analysis, wang2021cryogenic}:

\begin{equation}
    S_{21}(\omega) = 1 - \frac{2\frac{Q_l}{Q_c} e^{i \phi}}{1 + 2iQ_l \frac{\delta \omega}{\omega}} 
\end{equation}
where $\omega = 2\pi f$ is the angular frequency, $Q_l$ is the loaded quality factor, $Q_c$ is the coupling quality factor, $\phi$ is a phase that accounts for imperfections (e.g. reflections due to imperfect impedance matching) in the resonator coupling and $\delta \omega = \omega - \omega_0$ is the offset from the resonator frequency $\omega_0$. The internal quality factor is then extracted by the diameter correction method \cite{khalil2012analysis} as:

\begin{equation}
    \frac{1}{Q_i} = \frac{1}{Q_l} - \frac{\cos(\phi)}{Q_c}
\end{equation}

Instead of quality factor $Q_i$, we report the equivalent decay rate defined as $\kappa = \frac{f}{2 Q_i}$ in the main text. We subtract the zero field decay rate when reporting the decay rate as a function of temperature. The zero field decay rate is an intrinsic property of the resonator and does not arise from spin resonance.  We use a non-linear least squares fitting numerical package $\texttt{lmfit}$ to fit both real and imaginary components of the measured $S_{21}$ \cite{newville2016lmfit}. A further median filter with a 3-5 point window is applied to the decay rate vs magnetic field to remove single point jumps (likely due to superconducting vortices) from the dataset.

Note that the information about various loss mechanisms for photons in the resonator is contained in the internal quality factor $Q_i$. These loss mechanisms include a finite resistive impedance of the superconductor at temperatures close to $T_c$, electic-dipole coupled two-level systems in the surface oxide of the superconductor or at superconductor-substrate interfaces, superconducting vortices and magnetic-dipole coupled spins. Out of all these loss mechanisms, only the spins are expected to have to response that requires tuning into a resonance condition ($g \mu_B B = h f$) as described in the main text. As a result, by tuning the magnetic field and extracting the decay rate, we are able to separate out the spin-resonance contribution from other non-resonant loss mechanisms.  

When the magnetic field is swept, the resonator frequency discontinuously jumps at certain non-repeatable magnetic fields potentially due to introduction of vortices in the resonator CPW or the surrounding ground plane. In order to mitigate this problem, we use a automatic peak-finding algorithm that does a broad frequency scan ($50\,\si{\MHz}$ range) around the expected resonator frequency to find the resonator dip and then does a finer scan ($3\,\si{\MHz}$ range) to accurately measure the resonator profile. This peak-finding algorithm is performed at each magnetic field and temperature point in our dataset. Additionally, sweeping the magnetic field from higher fields to zero causes a reduction in the number of these jumps. A combination of such measurement strategies enables us to accurately and efficiently extract the resonator properties as a function of magnetic field and temperature without confounding background signals.    
\clearpage

\appendix
\title{Supplementary Information :  Microwave spin resonance in epitaxial thin films of spin liquid candidate $\text{TbInO}_3$}
\maketitle
\beginsupplement
\begin{bibunit}
\onecolumngrid

\section{Deposition of Superconducting films}
NbTiN films are deposited by magnetron sputtering of Nb and Ti targets in a Ar/N$_2$ environment. The recipe used for all resonators described in this work is as follows:

\begin{table}[h]
    \centering
    \begin{tabular}{|c|c|}
    \hline 
    Deposition Pressure & $15\,\text{mTorr}$  \\ \hline
    Ti source power & $200\,\si{\watt}$  \\ \hline
    Nb source power & $150\,\si{\watt}$  \\ \hline
    Substrate bias & $100\,\si{\watt}$  \\ \hline
    N$_2$/Ar flow ratio & $7.5\,\%$  \\ \hline
    Deposition temperature & $20\,\si{\celsius}$  \\ \hline
     Deposition time & $1200\,\si{\second}$  \\ \hline       
    \end{tabular}
    \caption{Parameters used for sputtering of NbTiN films}
    \label{table-sputter}
\end{table}
The NbTiN thickness as determined by atomic force microscope (AFM) imaging was $48\,\si{\nano\meter}$. The deposition recipe was optimized for the resonator quality Q factor measured at cryostat base temperature to exceed $0.5-1\times 10^6$ on Si substrates. The same recipe was then used for deposition on TbIn$O_3$ films as well as bare YSZ substrates. While it would be preferable to optimize relevant superconducting properties directly on the film of interest (e.g. TbInO$_3$/YSZ), we decided to replicate the same recipe independently optimized on Si for simplicity. Future work could focus on carefully optimized epitaxial growth of superconductors directly on spin liquid films of interest.  

\clearpage
\section{Device design and optical images of resonators on TbInO$_3$/YSZ}
\begin{figure*}[h]
    \centering
    \includegraphics[width=0.7\textwidth]{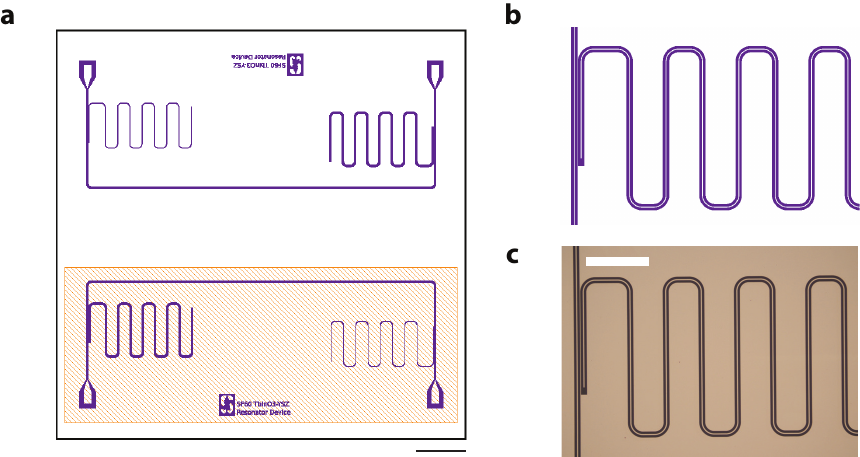}
    \caption{(a) Design file for the resonator device used in the main text. The black outer square denotes the boundary of a $10\,\si{\milli\meter}\times10\,\si{\milli\meter}$ chip. The region denote by the orange hatched rectangle is ion-milled to remove the TbInO$_3$ film as grown on the YSZ substrate. Two microwave feedlines with identical resonators are patterned on the same chip. The top device is used for TbInO$_3$/YSZ measurements while the bottom device is used to determine the YSZ substrate contribution.  Black scale bar at bottom right is $1000\,\si{\micro\meter}$.(b) Etch mask for a CPW resonator coupled to a microwave feedline on the left. The region in dark blue is etched using reactive-ion etching. (c) Optical image of a CPW resonator used for measurements described in the main text. The superconducting NbTiN film has a golden-brown color. White scale bar is $500\,\si{\micro\meter}$.}
    \label{fig-s-cad}
\end{figure*}

The CAD design file for the resonator device capacitively coupled to a microwave feedline is shown in Fig.~\ref{fig-s-cad}. The width of the center pin and the spacing from the ground plane are chosen to maintain a $50\,\si{\ohm}$ impedance for the CPW. We assumed a dielectric constant of 27 for the YSZ substrate \cite{lanagan1989dielectric}.

\clearpage
\section{Crystal structure of TbInO$_3$}
\begin{figure*}[h]
\centering
\includegraphics[width=\textwidth]{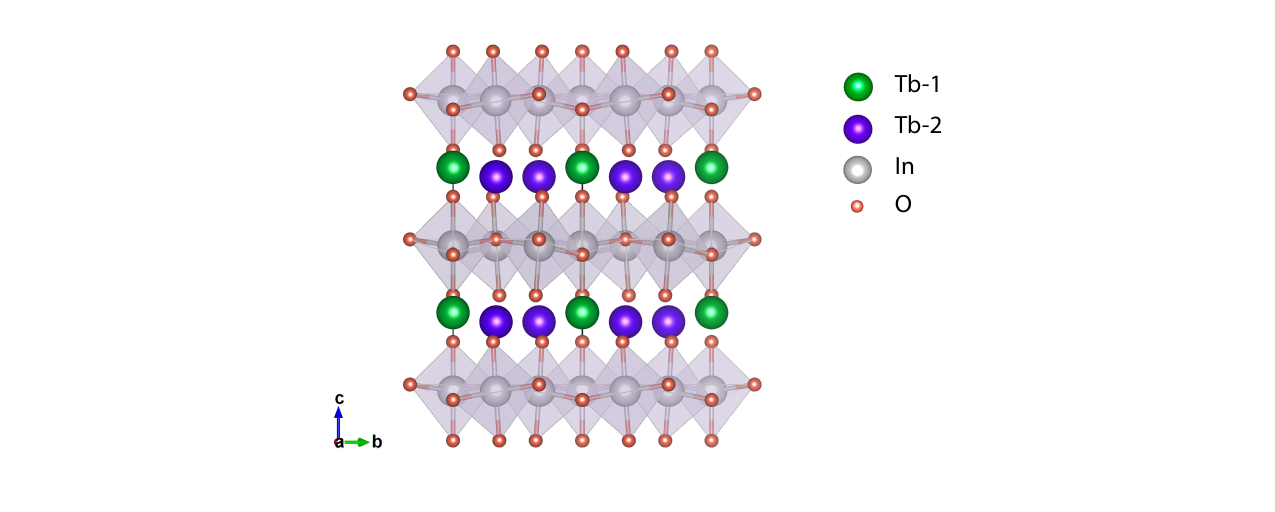}
\caption{Crystal structure of hexagonal TbInO$_3$ in the improper ferroelectric phase with P$6_3cm$ crystal symmetry. The structure consists of layers of magnetic Tb sites separated by layers of non-magnetic [InO$_5$]$^{7-}$ trigonal bipyramids. Due to the ferroelectric distortion, the Tb sites are split into two flavors, marked as Tb-1 and Tb-2 in the figure.}
\end{figure*}

\clearpage
\section{Microwave sample holder}

\begin{figure*}[h]
    \centering
    \includegraphics[width=0.8\textwidth]{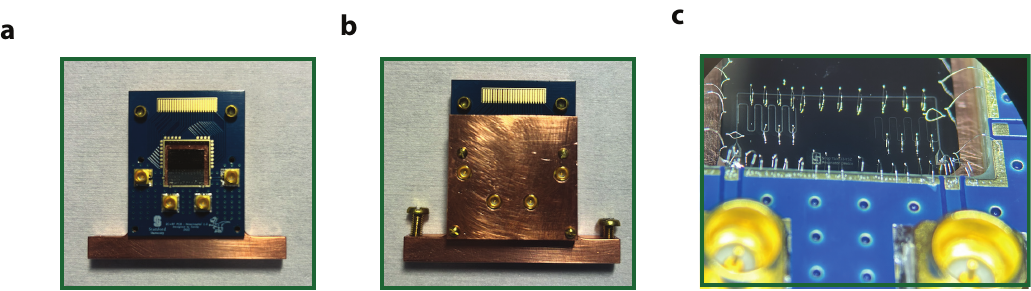}
    \caption{Microwave sample holder (a) A printed circuit board (PCB) with 4 mini-SMP connectors is rigidly anchored on a sample holder made of oxygen-free high conductivity (OFHC) copper. The chip with fabricated resonators is PMMA glued on the copper surface of the sample holder and is wire-bonded to CPW traces that go to the mini-SMP connectors. (b) A Cu lid placed screwed on top of the sample board rejects any microwave box-modes that could potentially couple to the resonator and degrade its quality factor. This entire assembly to rigidly anchored to a gold-plated Cu plate on a top loading probe which makes a thermal contact with the mixing chamber of the cryostat. (c) A zoomed-in view of the wire-bonded sample. We make multiple bonds from the PCB to the sample to minimize bondwire inductance. Bonds are also placed across the microwave feedline to connect ground planes to prevent parasitic slot-line modes. Mini-SMP connectors which connect to rest of the microwave setup on the cryostat are seen at the bottom edge of the image. }
    \label{fig-s-sample}
\end{figure*}

\clearpage
\section{Temperature dependence of Multi-Gaussian Fits}

\begin{figure*}[h]
    \centering
     \includegraphics[width=\linewidth]{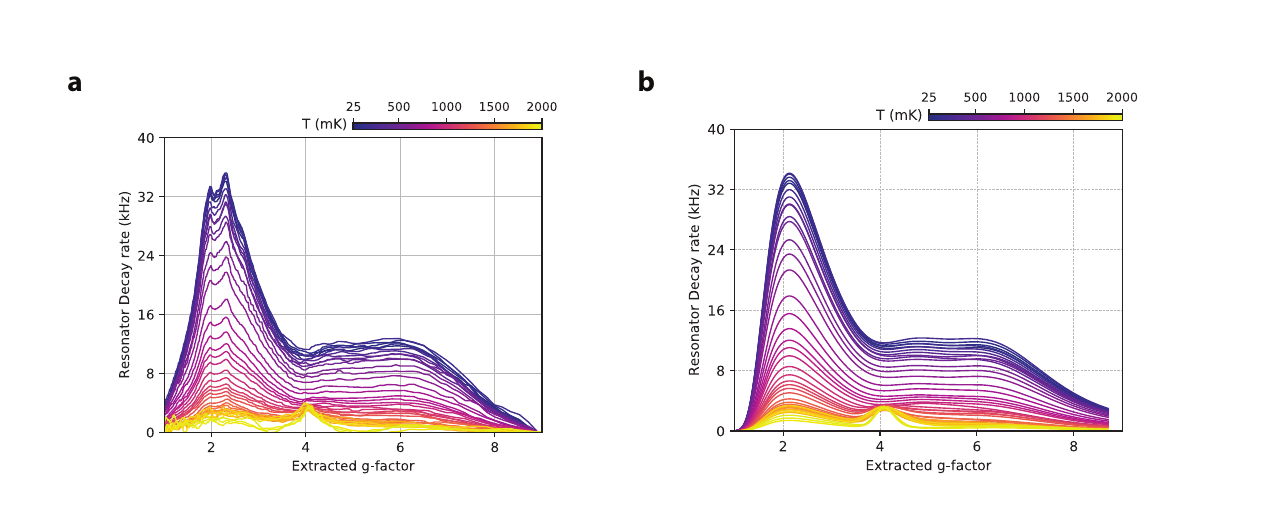}
    \caption{(a) Experimentally measured decay rate from a resonator on a TbInO$_3$ film across 20\,\si{\milli\kelvin} to 2\,K. A multi-peak Gaussian model is used to fit the decay-rate vs magnetic field data as a function of temperature. (b) The decay-rate vs g-factor as fit by the Gaussian model with varying temperature. Good agreement between the experimental data in (a) and the Gaussian model in (b) is seen. }
    \label{fig-s-multi-gauss}
\end{figure*}

\clearpage
\section{Device on a high-resistivity Si substrate}
\begin{figure*}[h]
    \centering
     \includegraphics{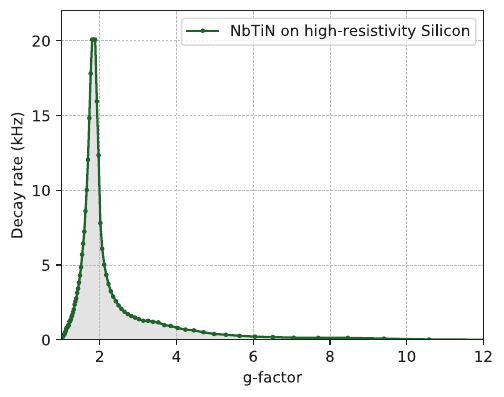}
    \caption{Spin resonance data from a resonator device fabricated on high-resistivity silicon. A single dominant resonance centered at $g=1.84 \pm 0.04$ is observed. }
    \label{fig-s-si}
\end{figure*}

In order to understand any background signals coming from the NbTiN superconductor, we fabricated similar resonator devices on high-resistivity silicon. The high-resistivity Si ensures that typical donor or acceptor atoms in Si which could be magnetic and have a spin-resonance response are absent. Fig.~\ref{fig-s-si} shows the decay rate plotted as a function of extracted g-factor from such a device. We observe a single dominant resonance centered at $g=1.84 \pm 0.04$. A similar g-factor has been recently reported on NbTiN resonators fabricated on sapphire substrates \cite{bahr2024improving}. It likely arises from paramagnetic defect states present in the surface oxide of NbTiN. While Nb$_2$O$_5$ with Nb$^{5+}$ oxidation state has no unpaired spins, any sub-oxide with a lower oxidation state has the potential to be a paramagnetic center. The absence of any other resonances in this device on a silicon substrate also indicates that the other resonances seen in the main text must arise from the TbInO$_3$ film or the YSZ substrate. 

\clearpage

\clearpage
\section{Data from additional TbInO$_3$ and YSZ devices}
\label{sec-chi}
\begin{figure*}[h]
    \centering
     \includegraphics[width=\linewidth]{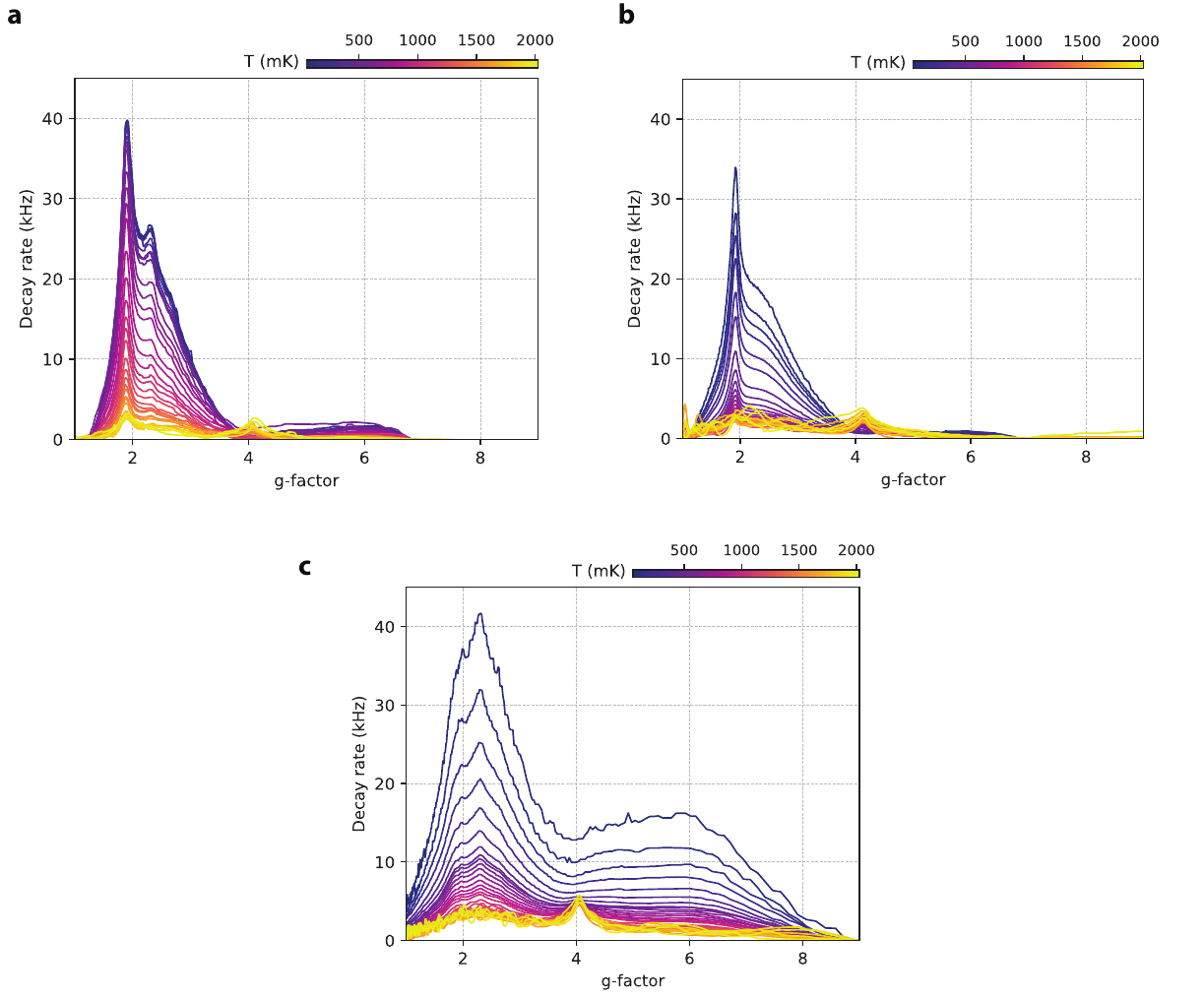}
    \caption{Spin-resonance data from three additional resonator devices: (a) Data from a resonator fabricated on the same chip as the main device in a region where TbInO$_3$ was removed using ion-milling (b) Data from a device on unprocessed YSZ substrate. (c) Data from a second TbInO$_3$/YSZ device.}
    \label{fig-s-adddata}
\end{figure*}

Fig.~\ref{fig-s-adddata} shows decay rate vs g-factor data from three additional resonator devices. Fig.~\ref{fig-s-adddata}a shows the spin resonance signal from a second resonator device fabricated on the same chip as the device in the main text. In addition to narrow $g=1.9$, $g = 2.3$ and $g=4.1$ impurity/defect lines as discussed before, we also observe a broad resonance centered at $g=2.1$. Hence, it is likely that broad $g=2.1$ signal (resonance R1 in the main text) seen when the TbInO$_3$ film is present arises either from the bulk YSZ substrate or from inter-facial defects at TbInO$_3$/YSZ interface. Fig.~\ref{fig-s-adddata}b shows the response from a resonator fabricated on an unprocessed YSZ substrate. Notably, the narrow line at $g=2.3$ is absent. This suggests an inter-facial origin to this defect state, possibly from transfer of oxygen atoms from the YSZ to the TbIn$O_3$ film as discussed in the main text. Fig.~\ref{fig-s-adddata}c shows data from a second device on a TbInO$_3$/YSZ film grown in a separate run from the device in the main text. With this device, we can quantitatively reproduce the measurements shown in fig. 2(c) on a different TbInO$_3$ film. The strength of the resonant features in terms of decay rate is slightly different ($\sim 10\,\%)$ which can possibly arise in slight variations of film thickness of the TbIn$O_3$ film. The location in terms of g-factors is unchanged as g-factors depend only on the local environment of the spins. 

\clearpage
\section{Estimation of magnetic susceptibility from area under ESR curve}
The area under the spin-resonance curve as a function of magnetic field is directly proportional to the DC susceptibility. This relationship is often used to extract a value in arbitrary units for the magnetic susceptibility from ESR measurements \cite{white1983quantum}. In order to make a comparison to traditional susceptibility measurements, we estimate the scaling factor from arbitrary units to emu mol$^{-1}$ G$^{-1}$, units traditionally used to report DC susceptibility measurements. 

Let $\chi(\omega) = \chi'(\omega) + i \chi''(\omega)$ denote the frequency dependent susceptibility of the spin system. For a magnetic field strength $H(\omega) = H_1 e^{i \omega t}$, the induced magnetization is given as,
\begin{equation}
    M (\omega) = \chi(\omega) H (\omega) = \chi(\omega) H_1 e^{i \omega t}
\end{equation}

The imaginary part of susceptibility represents a transfer (loss) of power from the magnetic field into the spin-system, which is them dissipated by the spins as they relax into the environment e.g. via phonons. The power absorption $P_1(\omega)$ is given as,

\begin{align}
    P_{1}(\omega, \vec{r}) &= -\Im{\left\langle \frac{d}{dt} (H^{*} M) \right\rangle } \\
    &= \frac{1}{2} \omega H_1^2 \chi" (\omega)
\end{align}

The total power lost by a resonator is given by the spatial integral of $P_1(\omega, \vec{r})$ over the mode volume of the resonator. Such a integral is required since the power absorption $P_1(\omega, \vec{r})$ is a function of the local magnetic field $H_1$ which is spatially non-uniform and depends on the resonator geometry. Let $P(\omega)$ denote the net power absorption over the resonator mode volume.

\begin{align}\label{eqn-p-abs}
    P(\omega) &= \int P_1(\omega, \vec{r}) n_s(\vec{r}) d^3 r \\
    &= \frac{1}{2} \omega \chi"(\omega) \int H_1^2(\vec{r}) n_s (\vec{r}) d^3 r
\end{align}
where $n_s(\vec{r})$ denotes the spin density. Assume that the spin density $n_s(\vec{r}) = n_s$ is spatially constant in regions where the spins are present and is zero otherwise. We introduce a participation ratio $\eta$ defined as follows:
\begin{align}
    \eta &= \frac{\int H_1^2(\vec{r}) \mathbb{I}_s(\vec{r}) d^3 r}{\int H_1^2(\vec{r}) d^3 r}
\end{align}
$\mathbb{I}_s(\vec{r})$ is an indicator function which is 1 in regions where the spin density is non-zero and zero otherwise.

For a simple harmonic oscillator, the denominator of the above expression is related to the average energy in the harmonic oscillator: $\frac{1}{2} \int H_1^2 d^3 r = \hbar \omega (\bar{n} + 1/2) \approx \hbar \omega \bar{n}$ where $\bar{n}$ is the average photon number in the resonator. P($\omega)$, the rate at which energy is absorbed by the spins and hence lost by the resonator is directly proportional to the decay rate $\kappa$ introduced in the main text. The proportionality factor is the average energy stored in the resonator. Therefore, we have $P(\omega) = \hbar \omega \bar{n} \kappa$. We can then simplify the expression \ref{eqn-p-abs} to get a relation between the experimentally measured resonator decay rate $\kappa$ and the imaginary component of susceptibility $\chi"(\omega)$.

\begin{align}
    \kappa(\omega) = \frac{1}{2} \omega n_s \eta \chi"(\omega) 
\end{align}

Finally, we use one of the Kramers-Kronig relations to equate the area under the imagining component of susceptibility, $\chi"(\omega)$ to the real component of DC susceptibility, $\chi'(0)$.

\begin{align}
    \chi'(0) &= \frac{1}{\pi} \int_0^{\infty} \frac{\chi''(\omega')}{\omega'} d\omega' \\
    &= \frac{2}{\pi n_s \eta}\int_0^\infty \frac{\kappa(\omega')}{\omega'^2} d\omega' 
\end{align}
We now change the integration variable from frequency $\omega$ to magnetic field $B$ using the relation $\omega = \gamma B = g \mu_B B$. $\gamma = g \mu_B$ is the gyromagnetic ratio.
\begin{equation}
    \chi'(0) = \frac{2}{\pi n_s \eta \gamma} \int_0^\infty \frac{\kappa(B)}{B^2} dB
\end{equation}

Finally, we convert to molar DC susceptibility $\chi_m$ with units of emu Oe$^{-1}$ mol$^{-1}$. 

\begin{equation}\label{eqn-chi}
    \chi'(0) = \frac{2 N_A}{\pi n_s \eta \gamma} \int_0^\infty \frac{\kappa(B)}{B^2} dB
\end{equation}
Here, $N_A$ is the Avogadro number. Spin density is the density of Tb ions: $n_s = \frac{2}{\sqrt{3} a^2 c} = 2.35\times10^{21}$ cm$^{-3}$. $a = 6.319\,\r{A}$ and $c = 12.312\,\r{A}$ are the unit cell constants for TbInO$_3$.  We calculate the participation ratio $\eta \approx 5.7\times10^{-4}$ by numerically integrating the magnetic field of a coplanar waveguide \cite{murray2018analytical}. 

\clearpage
\section{Estimation of Fe impurity concentration}
In this section, we estimate the concentration of Fe impurities that give rise to the narrow peak at $g=4.1-4.3$ as discussed in the main text. Using the same line of reasoning as section~\ref{sec-chi}, we derive an expression for the volume susceptibility similar to eq.~\ref{eqn-chi}.

\begin{equation}
    \chi'_v(0) = \frac{2}{\pi \eta \gamma} \int_0^\infty \frac{\kappa(B)}{B^2} dB
\end{equation}

Assuming these impurities are distributed uniformly throughout the bulk of the substrate, we assume a participation ratio of $\eta=0.5$, as half of the mode volume of the resonator is coupled to the substrate. The other half above the substrate is vacuum. Using data at 2K and integrating only around the $g=4.1-4.3$ peak, we get $\chi_v \approx 1.5\times 10^{-6}$ in CGS units. The Curie formula for volume susceptibility \cite{mugiraneza2022tutorial} is as follows: 

\begin{equation}
    \chi_v = \frac{n \mu_B^2}{k_B T}
\end{equation}
where $n$ is the density of spins and $T$ is temperature. For $\chi_v \approx 1.5\times 10^{-6}$, we obtain the impurity density $n \approx 4.8 \times 10^{18}\,\si{\centi\meter^{-3}}$. For simplicity, we assume the overall atomic number density of the YSZ substrate is $n_{\text{ZrO}_2} = 8.3\times10^{22}\,\si{\centi\meter^{-3}}$, the same as that of ZrO$_2$. Therefore, the impurity concentration is about $6\times 10^{-5} = 0.006\%$.

We can also estimate the density of Fe impurities by considering the ratio of susceptibility between Tb moments and impurity moments. Let $I_{\text{Tb}}$ be the area under the decay rate vs magnetic field curve for Tb moments and let $I_{\text{Fe}}$ be the corresponding area for Fe moments. Assuming the area is directly proportional to the product of the susceptibility and the participation ratio of the moments in the resonator, we get
\begin{align}
    \frac{I_{\text{Tb}}}{I_{\text{Fe}}} &= \frac{\chi_{\text{Tb}}}{\chi_{\text{Fe}}} \frac{\eta_{\text{Tb}}}{\eta_{\text{Fe}}} \\
    &= \frac{n_{\text{Tb}}}{n_{\text{Fe}}}\frac{\mu^2_{\text{Tb}}}{\mu^2_{\text{Fe}}} \frac{T_{\text{Fe}} - \theta_{\text{CW,Fe}}}{T_{\text{Tb}} - \theta_{\text{CW,Tb}}}\frac{\eta_{\text{Tb}}}{\eta_{\text{Fe}}}
\end{align}
Therefore, we can solve for $n_{\text{Fe}}$ in terms of $n_{\text{Tb}}$ as follows:
\begin{equation}
    n_{\text{Fe}} = \frac{I_{\text{Fe}}}{I_{\text{Tb}}}\frac{\mu^2_{\text{Tb}}}{\mu^2_{\text{Fe}}} \frac{T_{\text{Fe}} - \theta_{\text{CW,Fe}}}{T_{\text{Tb}} - \theta_{\text{CW,Tb}}}\frac{\eta_{\text{Tb}}}{\eta_{\text{Fe}}}
\end{equation}

We use $g=6.6$ for sites described by resonance R3 in the main text. For Fe impurities, we use $g=4.1$. Since the Fe impurities order below 800\,\si{\milli\kelvin}, we use the data at 2\,\si{\kelvin}. For Tb, we use data at base temperature of 50\,\si{\milli\kelvin} and $\theta^{LT}_{CW} \approx = -86\,\si{\milli\kelvin}$ as discussed in the main text.  Using the appropriate values for the participation ratios ($\eta_{\text{Tb}} = 5.7 \times 10^{-4}$ and $\eta_{\text{Fe}} = 0.5$) as discussed before, we get $n_{\text{Fe}} \approx 0.03 n_{\text{Tb, R3}} $. Note that $n_{\text{Tb, R3}}$ here is only the density from one flavor of the Tb sites, making up one third of all the Tb sites. Using the total density $n_{\text{Tb}} = 2.35\times10^{21}$ cm$^{-3}$ of Tb sites in TbInO$_3$, we get $n_{\text{Fe}} = 7.0\times 10^{19} \,\si{\centi\meter^{-3}}$.

We again use $n_{\text{ZrO}_2} = 8.3\times10^{22}\,\si{\centi\meter^{-3}}$ as the atomic number density of the YSZ substrate. Therefore, we get an impurity concentration of about $84\times 10^{-5} = 0.08\%$. Note that our two estimates differ by an order of magnitude with the discrepancy possibly arising from simplifying assumptions about the Curie-Weiss law made in those two estimates. Fe is a known impurity in YSZ, and either of our two estimates for Fe impurity concentration is plausible for these substrates.

\clearpage

\section{Frustration indices of spin liquid candidates}
The frustration index $f = |\theta_{\mathrm{CW}}|/T_{\mathrm{order}}$ is a commonly used metric for quantifying the degree of magnetic frustration. A value of $f \gtrsim 5$--$10$ generally indicates strong frustration~\cite{Ramirez1994}. In the following table, we list well-known spin liquid candidates with some of the largest frustration indices as reported in literature.

\begin{table*}[h]
\centering
\caption{Survey of frustration indices in geometrically frustrated magnets, sorted by descending frustration parameter $f = |\theta_{\mathrm{CW}}|/T_{\mathrm{order}}$. A value of $f \gtrsim 5$--$10$ generally indicates strong geometric frustration~\cite{Ramirez1994}. For materials with no long-range order (LRO) down to the lowest measured temperature, $f$ is given as a lower bound. SG denotes a spin-glass-like transition.}
\label{tab:frustration}
\renewcommand{\arraystretch}{1.25}
\begin{tabular}{l l c r r r l}
\hline\hline
Material & Lattice & $S$ & $\theta_{\mathrm{CW}}$ (K) & $T_{\mathrm{order}}$ (K) & $f$ & Refs.\ \\
\hline
$\kappa$-(BEDT-TTF)$_2$Cu$_2$(CN)$_3$ & Triangular & $1/2$ & $-375$ & ${\sim}0.032$ (no LRO) & $>10\,000$ & \cite{Shimizu2003} \\
ZnCu$_3$(OH)$_6$Cl$_2$ & Kagom\'{e} & $1/2$ & $-300$ & ${<}0.05$ (no LRO) & $>6000$ & \cite{Shores2005,Mendels2010} \\
NiGa$_2$S$_4$ & Triangular & $1$ & $-80$ & ${\sim}0.08$ (no LRO) & ${\sim}1000$ & \cite{Nakatsuji2005} \\
Na$_4$Ir$_3$O$_8$ & Hyperkagom\'{e} & $1/2$ & $-650$ & ${\sim}0.8$ (SG) & ${\sim}810$ & \cite{Okamoto2007} \\
Tb$_2$Ti$_2$O$_7$ & Pyrochlore & eff. & $-19$ & ${<}0.05$ (no LRO) & $>380$ & \cite{Gardner1999} \\
Ba$_3$CuSb$_2$O$_9$ & Triangular & $1/2$ & $-55$ & ${\sim}0.2$ (no LRO) & ${\sim}275$ & \cite{Zhou2011} \\
TbInO$_3$ & Triangular / honeycomb & eff. & $-17.2$ & ${<}0.1$ (no LRO) & $>170$ & \cite{Clark2019,Nordlander2025} \\
SrCr$_{9p}$Ga$_{12-9p}$O$_{19}$ (SCGO) & Kagom\'{e} (quasi-2D) & $3/2$ & $-500$ & ${\sim}3.5$ (SG) & ${\sim}143$ & \cite{Ramirez1990} \\
Ca$_{10}$Cr$_7$O$_{28}$ & Kagom\'{e} (bilayer) & $1/2$ & $-24$ & ${<}0.3$ (no LRO) & $>80$ & \cite{Balz2016} \\
Ba$_2$YMoO$_6$ & FCC (double perovskite) & $1/2$ & $-160$ & ${\sim}2.1$ (no LRO) & ${\sim}76$ & \cite{deVries2010} \\
\hline\hline
\end{tabular}
\end{table*}

\putbib[si-refs]
\end{bibunit}

\end{document}